\documentclass{amsart}

\usepackage{amsmath}
\usepackage{paralist}
\usepackage{graphics} 
\usepackage{epsfig} 
\usepackage{graphicx}
\usepackage{epstopdf}
\usepackage{subcaption}

\title[Models, measurement and inference] 
{Models, measurement and inference in epithelial tissue dynamics}

\author[O.J. Maclaren, A.G. Fletcher, H.M. Byrne and P.K. Maini]{}

\subjclass{Primary: 92CXX; Secondary: 92BXX, 92D25}
\keywords{epithelia, cancer, compartmental modelling, individual-based modelling, multiscale modelling, continuum mechanics, inference}

\email{maclaren@maths.ox.ac.uk}
\email{fletcher@maths.ox.ac.uk}
\email{byrneh@maths.ox.ac.uk}
\email{maini@maths.ox.ac.uk}

\begin{document}
\maketitle

\centerline{\scshape Oliver J. Maclaren}
\medskip
{\footnotesize
\centerline{Wolfson Centre for Mathematical Biology, Mathematical Institute, University of Oxford,}
\centerline{Radcliffe Observatory Quarter, Woodstock Road, Oxford, OX2 6GG, UK}
}

\medskip

\centerline{\scshape Helen M. Byrne}
\medskip
{\footnotesize
\centerline{Department of Computer Science, University of Oxford,}
\centerline{Wolfson Building, Parks Road, Oxford OX1 3QD, UK}}

\medskip

\centerline{\scshape Alexander G. Fletcher and Philip K. Maini}
\medskip
{\footnotesize
\centerline{Wolfson Centre for Mathematical Biology, Mathematical Institute, University of Oxford,}
\centerline{Radcliffe Observatory Quarter, Woodstock Road, Oxford, OX2 6GG, UK}}

\bigskip


\begin{abstract}
The majority of solid tumours arise in epithelia and therefore much research effort has gone into investigating the growth, renewal and regulation of these tissues. Here we review different mathematical and computational approaches that have been used to model epithelia. We compare different models and describe future challenges that need to be overcome in order to fully exploit new data which present, for the first time, the real possibility for detailed model validation and comparison.
\end{abstract}


\section{Introduction}

There are many exciting challenges facing mathematical modelling in biology. 
In this article we review some of these challenges in the context of the healthy and diseased dynamics of a specific biological system, the epithelial tissues that line internal and external body surfaces. 
Much of the interest in epithelial tissues arises because almost all solid cancers are epithelial in origin \cite{wright1984biology}; however, many of the modelling issues we discuss are relevant to a wider range of problems in mathematical biology.

The challenges we will highlight are: (I) \textit{Scale and resolution of models}. The desire for a model to address a specific biological question often motivates model simplicity; however, it is now well-known that biological function arises from the integration of coupled processes acting across a range of spatial and temporal scales. Often these processes cannot be easily separated and this creates technical challenges for mathematical and computational modelling. (II) \textit{Model-model comparison}. Currently, many different models have been developed to address the same problem, leading us to ask under what conditions do the different models generate the same predictions? Further, when models disagree, which approach is better? Questions of this nature motivate the development of new methods of model-model comparison. (III) \textit{Model-data comparison}. Validation of multiscale models is typically performed in a coarse-grained manner. With the advent of improved imaging techniques, there is now the potential to compare such models with high-resolution experimental data. Modern statistical inference methods, made available with increased computing power \cite{berger2000bayesian}, have yet to be fully exploited, however, and offer exciting opportunities for the future.  

The remainder of the paper is structured as follows. Section~\ref{sec:epithelial_biology} contains a brief biological introduction to epithelial tissues. We take the view that many questions regarding epithelial pathology such as tumorigenesis can be profitably understood in the context of failure of normal homeostasis and repair mechanisms. Thus we consider a range of phenomena in epithelial tissue dynamics. In Sections~\ref{sec:compartment_models} and~\ref{sec:individual_based_models}, respectively, we review compartmental and individual-based approaches that have been used to model epithelial cell populations, emphasising how the scale and resolution of the models (challenge I) have yielded insights into epithelial dynamics, and how some organising concepts have emerged within these model classes. 
At the same time, the absence of a common structure linking the different compartmental and cell-based models makes their comparison (challenge II) difficult. This motivates our discussion in Section~\ref{sec:continuum_models} of continuum limits of discrete, cell-based models, considered in the context of mechanically-interacting epithelial cells. In the discussion in the final section, Section~\ref{sec:discussion}, we focus on the opportunities that new data raise for measurement and parameter estimation (challenge III), referencing examples from tumour dynamics and epithelial development and repair, while emphasising the intertwined nature of the three challenges I-III. We also speculate on possible approaches for developing unified frameworks for using modelling, measurement and inference to better understand epithelial tissue dynamics. 


\section{Epithelial biology} \label{sec:epithelial_biology}

\subsection{Motivation}

The epithelial tissues that line the internal and external surfaces of the body are a primary source of physical protection and act as barriers to unwarranted substances~\cite{marchiando2010epithelial}. 
Maintenance of normal physiological function requires that many epithelia simultaneously carry out directed transport of substances such as water and electrolytes between bathing solutions on either side of the epithelial tissue; this is crucial for tasks such as regulating body fluid composition and absorbing nutrients~\cite{reuss2011epithelial}. 
As a result, many epithelial tissues, such as the skin and intestinal epithelia, have rapid rates of self-renewal and distinctive structural and functional organisation (see below). 
Importantly, dysregulation of the renewal process is thought to be associated with the large number (up to 90\%) of cancers originating in epithelia~\cite{radtke2005self, wright1984biology}. 
Epithelial tissues thus serve as important biological model systems for studying the dynamics of multicellular populations; further, understanding them under both healthy and diseased conditions is important for understanding the transition from one state to the other.

\subsection{General structure and function}
\begin{figure}[h]
\begin{subfigure}{\textwidth}
\centering
\includegraphics[width=0.6\linewidth]{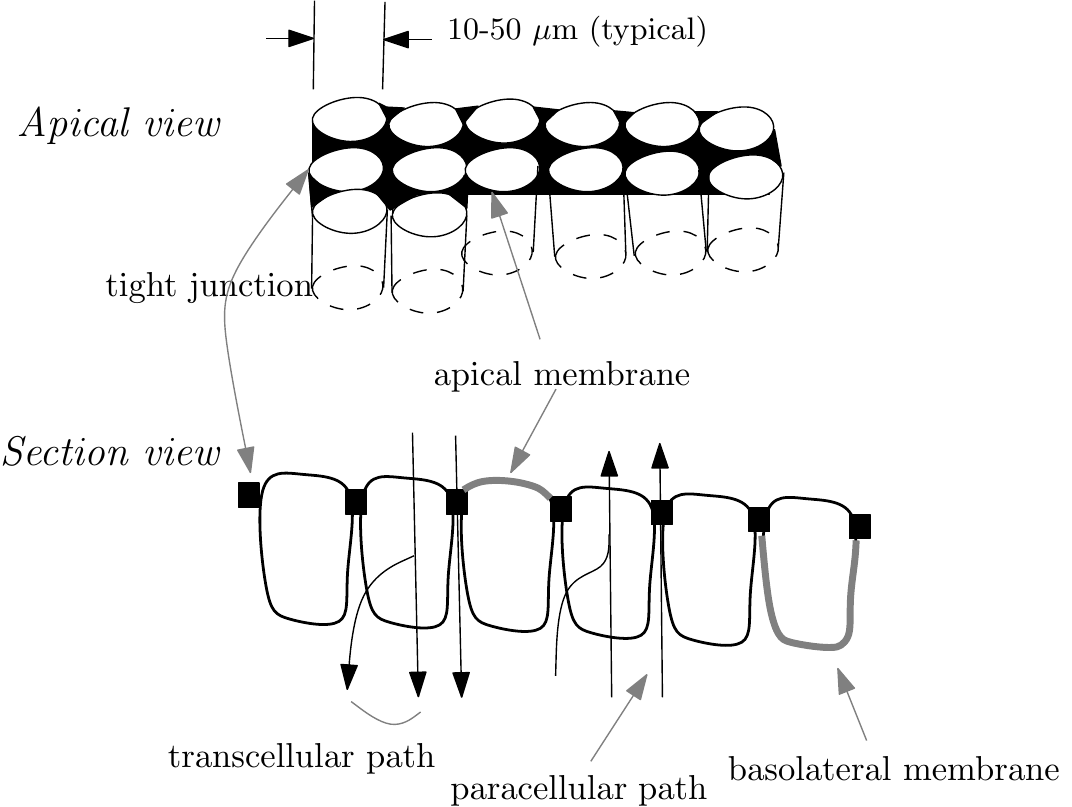}
\caption{}
\end{subfigure}
\begin{subfigure}{\textwidth}
\centering
\includegraphics[width=0.6\linewidth]{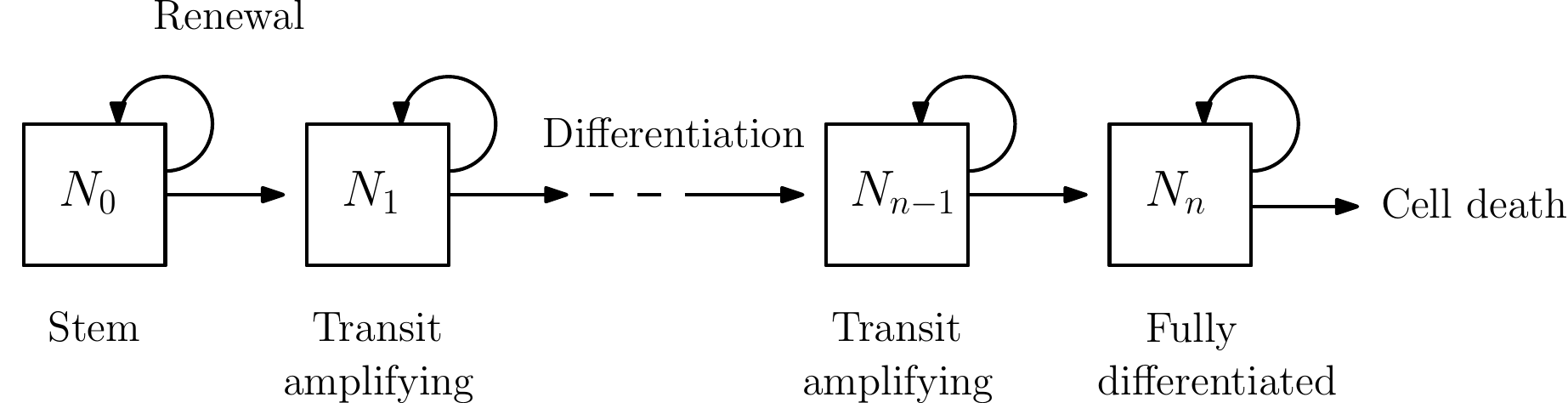}
\caption{}
\end{subfigure}
\begin{subfigure}{\textwidth}
\centering
\includegraphics[width=0.4\linewidth]{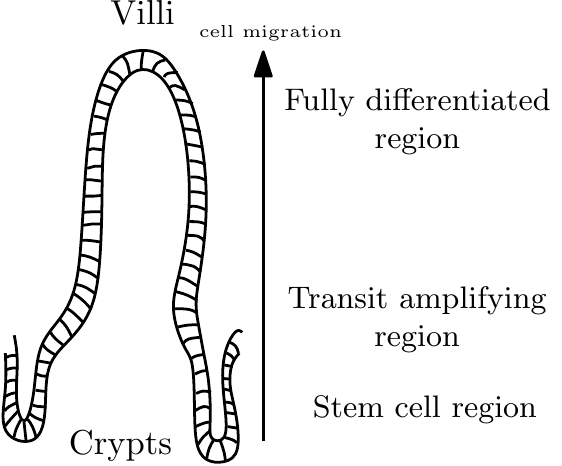}
\caption{}
\end{subfigure}
\caption{A: Schematic of the polarised structure of epithelial sheets, adapted from~\cite{friedman2008principles}. 
B: Compartment description of the lineage structure of epithelial cell populations. 
C: Schematic of the structure of the small intestine, showing crypt invaginations where stem and transit amplifying cells are located and villus protrusions, responsible to the absorption of nutrients and water. Based on~\cite{britton1982mathematical}.}
\label{fig:epi}
\end{figure}

Epithelial tissues comprise sheets of polarised cells, as well as glandular ingrowths (or crypts) and finger-like protrusions (or villi) formed from these sheets; 
the arrangement of these layers is maintained by specialised cell-cell and cell-matrix junctions~\cite{reuss2011epithelial} (Figure~\ref{fig:epi}A). These junctions, especially the tight junctions, are crucial not only for structural integrity of the epithelial layers, but also for maintaining cell polarity and, hence, the ability of epithelial tissues to perform directed transport of solutes and water. 
Individual epithelial cells are subject to strong mutual mechanical interactions mediated through these proteins. 

The self-renewal of epithelial tissues is driven by stem cells which generate, in a tightly-controlled manner, a sequence of increasingly specialised descendants known collectively as a lineage structure (Figure~\ref{fig:epi}B). 
This sequence must produce a sufficient number of descendants to maintain those parts of the epithelium exposed to the greatest levels of damage and so the spatial arrangement of the cell lineages frequently corresponds to functional requirements. 
An archetypal example is provided by intestinal crypts (Figure~\ref{fig:epi}C), finger-like invaginations of the intestinal mucosa. 
The intestinal crypt is structured spatially into regions corresponding to a stem cell niche at the base, an intermediate region of rapidly proliferating transit-amplifying cells, and a region of differentiated, functional cells nearest to the crypt orifice, where most transport and damage occur~\cite{wright1984biology}. 
This lineage structure maximises the number of functional descendants of stem cells while protecting the stem cells from damage. 
In addition to extensive experimental study, the intestinal epithelium has been the focus of numerous mathematical models, especially in the context of colorectal carcinogenesis~\cite{carulli2014unraveling, dematteis2013review, fletcher2010multiscale, johnston2007examples, kershaw2013colorectal}. 
In the sections that follow we will review some of these models. 

\section{Compartmental models} \label{sec:compartment_models}

\subsection{Overview}

Compartmental models provide a flexible framework for studying cell population dynamics. As a result, they
have been widely used to study epithelial tissue homeostasis, growth, regeneration and carcinogenesis, and the interplay between these dynamics.
These models account for processes such as cell proliferation, differentiation and death in generic terms, without specifying details of the mechanisms (e.g. signalling pathways) by which they are regulated.
 
The resolution of these models is coarse; compartments typically correspond to either spatial domains or distinct cell populations, with movement or other exchange between compartments. Using too many compartments is not typically a good idea - it can lead to parameters that are redundant or poorly constrained. In this case one might seek to reduce the number of compartments or, alternatively, use partial differential equation (PDE) or delay-differential equation (DDE) models, which replace the discrete parameters with some constrained (e.g. smooth) function or `lumping' of parameters. In what follows we restrict attention to deterministic compartment models comprising systems of ordinary differential equations (ODEs). Further details on more complex compartment models, including those that account for age-structure and stochastic effects, may be found in~\cite{carulli2014unraveling} and references therein. Later sections address alternative models for introducing finer-grained spatial resolution.


\subsection{Framework}

The general form of an ODE-based compartment model for a sequence of $n$ cell lineages (see Figure~\ref{fig:epi}B), which incorporates the processes of cell renewal, differentiation and death, is given by
\begin{align}
\label{eq:comp_models_start} \frac{dN_0}{dt} &= (\alpha_0-\beta_0-\gamma_0)N_0, \\
\frac{dN_1}{dt} &= (\alpha_1-\beta_1-\gamma_1)N_1+\beta_0 N_0,\\
\vdots \nonumber\\
\frac{dN_{n-1}}{dt}&= (\alpha_{n-1}-\beta_{n-1}-\gamma_{n-1})N_{n-1}+\beta_{n-2}N_{n-2},\\
\frac{dN_n}{dt}&= -\gamma_{n} N_{n} + \beta_{n-1}N_{n-1}.
\label{eq:comp_models_end}
\end{align}
Here $N_i (t)$ denotes the number of cells in compartment $i$ at time $t$, while the parameters $\alpha_i$, $\beta_i$ and $\gamma$ denote the renewal, differentiation and death rates of cells in compartment $i$. 
These parameters may be constants (for a given biological scenario) or may be prescribed functions of the different cell populations. In the latter case, the functional forms encode key hypotheses about how signalling and other control mechanisms regulate cell kinetics; we consider some examples of this next.


\subsection{Applications to homeostasis, control and tumorigenesis}

Johnston et al.~\cite{johnston2007mathematical} construct a model of the form~\eqref{eq:comp_models_start}--\eqref{eq:comp_models_end} to investigate homeostasis and the transition to tumorigenesis in the intestinal crypt. 
They consider three compartments, corresponding to stem, transit amplifying and differentiated cells respectively.
They note that models in which the kinetic parameters $\alpha_i$, $\beta_i$ and $\gamma_i$ are constants are {\em structurally unstable}: slightly different parameter values can lead to qualitatively different population dynamics. 
In their model, the total number of crypt cells may exhibit exponential growth, exponential decay or converge to a finite equilibrium, according to whether the net growth rate in the stem cell compartment, $\alpha = \alpha_0-\beta_0-\gamma_0$, is greater than, less than or equal to zero, respectively. 
Such behaviour is physically unrealistic, not least because the model parameters are idealisations, which are unlikely to remain constant during tissue homeostasis. 

To obtain a model in which homeostasis is preserved as system parameters vary, Johnston et al.~\cite{johnston2007mathematical} incorporate local feedbacks within the stem and transit amplifying compartments, such that the proportion of cells differentiating in a given compartment varies with the compartment size. 
They find that a feedback which increases linearly with compartment population size $N$, taking the form $\alpha(N) = a+bN$ for constants $a$ and $b$, leads to a model for which homeostasis is a stable steady state. 
While the total number of crypt cells at equilibrium changes as parameter values vary, representing the effect of pre-cancerous mutations on cell kinetics, the system does not exhibit the unbounded growth that is characteristic of tumorigenesis. Improving on this, the authors find that including a saturating local feedback of the form $\alpha^*(N) = a^* + b^*N/(1+c^*N)$, for constants $a^*$, $b^*$ and $c^*$, leads to a model that, as parameter values are varied to represent the effects of mutations, exhibits both changes to equilibrium cell numbers as well as eventual unbounded growth after sufficient mutations accumulate. The authors note that this behaviour recapitulates the natural pattern of tumorigenesis, where periods of rapid expansion alternate with long periods of equilibrium in a stepwise manner.


In other work, Lander et al. \cite{lander2009cell} (previously analysed  in more generality in \cite{lo2009feedback}) develop an ODE-based compartment model to study the renewal of mammalian olfactory neural epithelium. In contrast to the model of Johnston et al. \cite{johnston2007mathematical}, this model allows for feedback mechanisms whereby cells that are further along the lineage modulate the behaviour of cells earlier in the lineage, this feedback being mediated by chemicals sometimes given the generic name `chalones'. 

Lander et al. \cite{lander2009cell} consider the cases of two, three and an arbitrary number of compartments, and assume that cell death is restricted to the most differentiated compartment only: in all other compartments, dividing cells either renew or differentiate. Lander et al. denote by  $0 \leq p \leq 1$ the renewal probability and by $0 \leq v$ the rate of cell division, and so we may identify the link between our notation and theirs through $\alpha-\beta = (2p-1)v$.

Lander et al.~\cite{lander2009cell} use an engineering approach to analyse their model.
They consider a variety of control `objectives' such as control of size, growth rate and cell-type proportions. 
In agreement with Johnston et al.~\cite{johnston2007mathematical}, they observe that without feedback the system is sensitive to variations in parameter values. 
They consider several feedback mechanisms by which chalones may achieve the control objectives. These include feedback on renewal probabilities, division rates, and the compartments between which the feedbacks operate. 
Feedback on renewal probabilities, rather than division rates, is found to give the most robust form of control, with reduced parameter sensitivity, faster regeneration in response to injury and less sensitive dependence on initial conditions. 

The work of Lander et al.~\cite{lander2009cell} illustrates how even simple theoretical analyses can generate new biological insight when carefully combined with experiments. The authors progress from a model prediction, which was unexpected without formal analysis, to experimental validation by considering the protein growth differentiation factor 11 (GDF11), a member of the transforming growth factor beta superfamily. While already known as an inhibitor of the mammalian olfactory neural epithelium, GDF11's primary target was believed to be the rate of cell division rather than the renewal probability. 
Lander et al. show experimentally that GDF11 appears negatively to regulate both rates, and conclude that many experimental observations are consistent with the theoretical expectations of `best' control.


The generic nature of compartment models is both a strength and a weakness. 
Since they are formulated in phenomenological terms, they can be used to determine what types of feedback
mechanisms will produce robust dynamic behaviour when model parameters fluctuate, and how these might fail.
However, it can be difficult to relate these general features to specific measurable quantities and/or to rule out 
alternative mechanisms. 
Furthermore, while capable of representing coarse spatial structure, such as cell lineages in the intestinal crypt, compartmental models lack the spatial resolution needed to track cell movement or to account for mechanical effects that regulate such movement. This can have important implications for understanding both healthy and diseased state dynamics. 
We next consider a modelling approach that can address such issues.


\section{Individual-based models} \label{sec:individual_based_models}

\subsection{Overview}

\begin{figure}[hbt]
\begin{subfigure}{.5\textwidth}
\centering
\includegraphics[width=.9\linewidth]{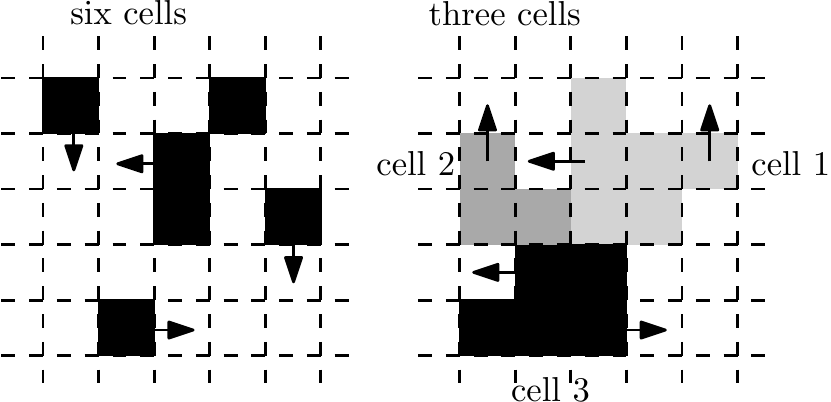}
\caption{}
\end{subfigure}
\begin{subfigure}{.5\textwidth}
\centering
\includegraphics[width=1.0\linewidth]{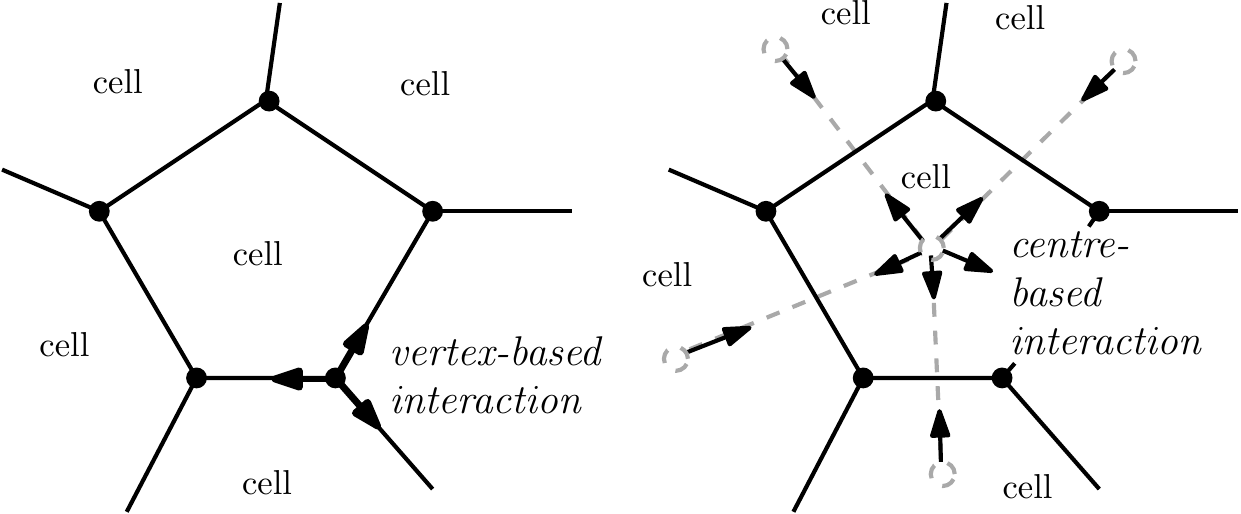}
\caption{}
\end{subfigure}
\begin{subfigure}{.5\textwidth}
\centering
\includegraphics[scale=0.7]{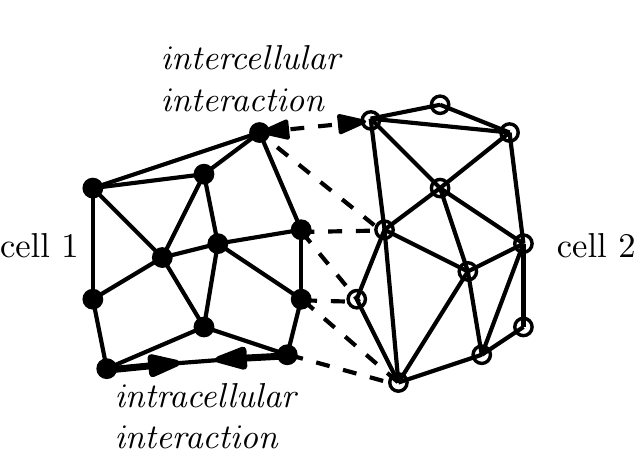}
\caption{}
\end{subfigure}
\caption{A: Schematic of a cellular automaton model (left), where each shaded lattice site corresponds to one cell, and a cellular Potts model (right), where each collection of similarly coloured lattice sites represents one distinct biological cell. 
B: Schematic of an off-lattice, vertex-based model (left), where the solid points are the cell vertices mediating the mechanical forces (indicated by the black arrows), and an off-lattice, cell centre-based model (right), where mechanical interactions (indicated by the black arrows) apply between cell centres.
C. Schematic of a subcellular element model with two cells, each represented by a collection of similarly coloured vertices called `subcellular elements', and where there are both intracellular (solid arrows) and intercellular (dashed arrows) interactions between elements.
Images based on those in~\cite{kershaw2013colorectal}, ~\cite{osborne2010hybrid} and \cite{newman2005modeling}.}
\label{fig:ibm}
\end{figure}

In order to account for phenomena such as cell sorting, rearrangements, mechanical interactions and pattern formation, mathematical models with spatial resolution are often required. 
Individual-based models (IBMs) offer a natural framework for describing the behaviour of a finite number of individual cells or cellular components. 
These models are typically termed either {\it `on lattice'} or {\it `off lattice'}, depending on how individual cells are represented (Figure~\ref{fig:ibm}). 
In the following we focus on general features of IBMs that have been used to study the dynamics of cell sorting in embryonic epithelia and the influence of mechanical forces on the dynamics of intestinal crypt cells: for more
extensive reviews, see~\cite{dematteis2013review} and~\cite{kershaw2013colorectal}.

\subsection{Applications to carcinogenesis, cell sorting and intestinal crypt dynamics}

\subsubsection*{On-lattice models}

On-lattice models usually discretise space using a fixed, regular, lattice. For example,
in cellular automata (CA), each cell occupies a single lattice site, which may change over 
time~\cite{dorman2005cellular}. 
The system dynamics are encoded in a simple, rule-based manner, with changes in cell configuration occurring at discrete time steps, using synchronous or asynchronous updating or an event-driven approach based on the Gillespie algorithm~\cite{block2007classifying, lee1995cellular}.
CA have no \textit{a priori} requirement to satisfy physical principles and hence may exhibit unphysical dynamics. 
For example, when a cell divides in a CA, its daughter cells may `shove' their neighbours to make space, resulting in instantaneous displacements over many cell diameters. 
Model simulations may also suffer from artificial anisotropies induced by symmetries of the underlying lattice. Despite their relative simplicity, CA models have been successfully used to predict aspects of epithelial carcinogenesis. For example, Gatenby et al. \cite{gatenby2007cellular} combined CA simulations with histological studies of tumour biopsies to demonstrate the key role that adaptation to hypoxia and acidosis may play in the transition from a precancerous lesion (such as ductal carcinoma in situ in the breast) to invasive cancer.

The cellular Potts model (CPM) is another popular, on-lattice-model, based on simple models of magnetism from statistical mechanics, and adapted by Graner and Grazier \cite{graner1992simulation, glazier1993simulation, glazier2007magnetization} to study cell sorting in embryonic epithelia. 
In the CPM, cells may span several, connected lattice sites, rather than being in one-to-one correspondence as for the CA models. 
If lattice sites are indexed by a single integer $i$, then cells correspond to collections of lattice sites which share the same value of an integer label $\sigma(i)$. Using $\mathcal{N}$ to denote the set of all pairs of neighbouring lattice sites $(i,j)$, the possible interactions between neighbouring cells are encoded via a Hamiltonian (interaction energy), such as \cite{graner1992simulation, glazier1993simulation}
\begin{align}
H = \sum_{(i,j) \in \mathcal{N}} \left(1-\delta_{\sigma(i)\sigma(j)} \right)J(\tau(\sigma(i)),\tau(\sigma(j)))+\lambda_A \sum_{\sigma}(A(\sigma)-A_T(\tau(\sigma)))^{2},
\label{eqn-cpm}
\end{align}
where the additional label $\tau$  defines cell type and $\delta_{mn}$ is the Kronecker delta. The first summation in equation (\ref{eqn-cpm}) represents an adhesion interaction energy and the second term an energy associated with cell stretching. In (\ref{eqn-cpm}), different adhesion energies $J(\tau(\sigma(i),\tau(\sigma(j))$ may be assigned to interactions involving different types of cells, but individual cells do not have adhesive interactions with themselves. The area of a cell (assuming we are modelling cells in two dimensions) is denoted by $A=A(\sigma)$, $A_T(\tau(\sigma))$ is its target area and $\lambda_A$ is an tunable phenomenological parameter weighting the contribution of the target area energy to the adhesive energy. Additional interaction or cell property terms may be added as desired. The model dynamics are implemented via a Markov chain Monte Carlo (MCMC) updating method (the Metropolis algorithm) rather than an explicit equation of motion \cite{graner1992simulation, glazier1993simulation}. We note that the original statistical mechanical models that the CPM is based on were developed to study equilibrium systems and there is some difficulty in interpreting the dynamic features of the CPM (see e.g. the discussion in \cite{vossbohme2012multi}).


As mentioned, the CPM was introduced by Glazier and Graner \cite{glazier1993simulation, graner1992simulation} to explain experiments performed on embryonic epithelial cell populations. 
After artificial mixing, the different types of cells rearrange until they recover their pre-mixed configurations. 
Steinberg~\cite{steinberg1962mechanismII, steinberg1962mechanismI, steinberg1962mechanismIII, steinberg1963reconstruction} introduced the differential adhesion hypothesis (DAH) to explain this phenomenon, the general principle being that intermixed cell populations rearrange so as to minimise their collective adhesive free energy. The CPM, as implemented by Graner and Glazier \cite{glazier1993simulation, graner1992simulation}, successfully, albeit qualitatively, reproduces observations of cell sorting~\cite{glazier1993simulation, graner1992simulation} based on the DAH. This concept has subsequently been applied to processes in development, wound repair, and tumorigenesis; for example, \cite{turner2002intercellular} used a CPM to investigate the influence of changes in cell-cell and cell-substrate adhesion on cancer invasion. Their model notably incorporates a description of cell proliferation in which the mitotic rate depends on the relative magnitudes of homotypic and heterotypic adhesion. Despite difficulties in precisely relating some aspects of this model framework to biological processes, the CPM remains an informative and popular framework~\cite{dematteis2013review, glazier2007magnetization, ouchi2003improving}.

\subsubsection*{Off-lattice models}

Off-lattice IBMs describe the dynamics of discrete, mechanically-interacting cell entities. 
At their simplest these entities are typically either cell-centres or cell vertices, while more complex models allow for additional resolution of the cell interior in terms of further `subcellular'-scale entities. The basic governing principles are the same in all cases, however. As for the CPM, these models can be formulated in terms of potential functions encoding the interaction energy between cells or cellular elements. In contrast to the CPM, here an explicit equation of motion is specified for the cellular entities.

For vertex models we consider cells indexed by $\sigma = 1,...,N$, as in the CPM, but now with vertices, rather than grid points, indexed by $i=1,...,V$ and with each vertex having an associated position vector $\mathbf{r}_i(t)$. Here cell vertices represent the tight junction proteins connecting cells. The free energy of each cell is then defined as a sum of contributions, such as adhesive energy $U^J$, target area (assuming for convenience that we are in two-dimensions) energy $U^A$ and target perimeter energy $U^P$, giving \cite{osborne2010hybrid,fletcher2013implementing}
\begin{align}
U &= \sum_{\sigma} \left(U_{\sigma}^J + U_{\sigma}^A + U_{\sigma}^P\right)\\
&= \sum_{\sigma} \left(\lambda_J\sum_{L\in \mathcal{L}(\sigma)}L^2+\lambda_A (A(\sigma)-A_T(\sigma))^2+\lambda_P (P(\sigma)-P_T(\sigma))^2\right).
\label{eqn-vertex}
\end{align}
Here $\mathcal{L}(\sigma)$ is the set of lengths $L$ of the lines, or `bonds', connecting the vertices of cell $\sigma$; the first term, summing over these lengths, then represents the total vertex bond energy of cell $\sigma$. $A(\sigma)$ and $P(\sigma)$ are the area and perimeter of cell $\sigma$, respectively, and $A_T(\sigma)$ and $P_T(\sigma)$ are its target area and perimeter, respectively. Finally, $\lambda_A$, $\lambda_P$ and $\lambda_L$, respectively, are phenomenological parameters weighting the contributions of the above energies. 

As indicated, each cell $\sigma$ has an associated set of vertices with position vectors $\mathbf{r} 
\in \mathcal{R}(\sigma)$ and each of the above cell properties are then defined by functions of these position vectors. This allows an equation of motion to be written for the coordinate vectors of the vertices, here an overdamped Newtonian equation. This is driven by (generalised) forces $\mathbf{F}_i$ obtained by differentiating the free energy function considered as a function of vertex coordinates. This gives
\begin{align}
\eta_{i} \frac{d\mathbf{r}_{i}}{dt} = \mathbf{F}_{i} = - \mathbf{\nabla}_{\mathbf{r}_{i}} U(\mathbf{r}_1,...,\mathbf{r}_i,...\mathbf{r}_V),
\label{eq:motion-vertex}
\end{align}
where $\eta_{i}$ is the viscous drag coefficient of vertex $i$ and $\mathbf{\nabla}_{\mathbf{r}_{i}}$ the gradient operator with respect to $\mathbf{r}_i$. 

While cell-centre models may be formulated in energetic terms, they are typically (but equivalently) formulated by applying force balances. Thus the equation governing the motion of $\mathbf{r}_i$, where $i$ now indexes cell centres rather than the cell vertices, is similar in form
to (\ref{eq:motion-vertex}):
\begin{align}
\eta_{i} \frac{d\mathbf{r}_{i}}{dt} = \sum_{j \neq i} \mathbf{F}_{ij},
\label{eq:motion-centre}
\end{align}
where $\mathbf{F}_{ij}=-\mathbf{F}_{ji}$ is the assumed equal and opposite force that cell centre $j$ exerts on cell centre $i$, an example being the linear spring law for which  $\mathbf{F}_{ij} = k(\vert \mathbf{r}_{i}-\mathbf{r}_{j}-S_{Tij}\vert)(\mathbf{r}_{i}-\mathbf{r}_{j})/(\vert\mathbf{r}_{i}-\mathbf{r}_{j}\vert)$, where $S_{Tij}$ is the target length of the spring connecting cell centres $i$ and $j$. In (\ref{eq:motion-centre}), the sum is over all cell centres $j$ in a well-defined neighbourhood of the cell with centre $i$ (e.g. by a Delaunay triangulation).

The `subcellular element' approach was introduced by Newman \cite{newman2005modeling} as a general framework for investigating the emergent behaviour of cells and tissues, including tumorigenesis; the mechanics of an epithelial sheet was chosen as an initial example but many other systems have been studied (see below). In this model, cells are defined by a finer set of vertices than just cell boundary or centre positions. In this case $i$ indexes a set of subcellular-scale vertices called `subcellular elements', which have a phenomenological correspondence to internal cellular architecture. The total potential is split into a set of interaction potentials which here are further divided into intracellular and intercellular components depending only on the distance between element positions. A noise term $\xi$ is often added (this may similarly be added to the above vertex or cell-centre models), and is typically chosen to be Gaussian with zero mean. Again, however, the principles are the same. Here the equation of motion for the position vector $\mathbf{r}_{i}$ of a subcellular element $i$, with associated cell $\sigma$, takes the form

\begin{align}
\eta_{i} \frac{d\mathbf{r}_{i}}{dt} &= - \mathbf{\nabla}_{\mathbf{r}_{i}}U+\xi \\
&= - \mathbf{\nabla}_{\mathbf{r}_{i}} \left(\sum_{\mathbf{r_j} \in \mathcal{R}(\sigma)/\lbrace{\mathbf{r}_i\rbrace}}U_{\mbox{intra}}(\vert\mathbf{r}_i-\mathbf{r}_j\vert)+
\sum_{\sigma' \neq \sigma}\sum_{\mathbf{r_j} \in \mathcal{R}(\sigma')}U_{\mbox{inter}}(\vert\mathbf{r}_i-\mathbf{r}_j\vert)\right) 
+ \xi,
\label{eq:motion-subcellular}
\end{align}
where the first sum of accounts for interactions between element $i$ of cell $\sigma$ and all other intracellular elements of the same cell, while the second sum governs interactions of element $i$ with each element $j$ of each other cell $\sigma'$. 

Though the above models are all similar in form, each comes with its own conceptual framework for specifying the particular interaction potentials (i.e constitutive modelling); thus detailed comparisons of all three approaches under different circumstances would be helpful. This will be greatly aided by the development and sharing of open source software; an example is Chaste \cite{mirams2013chaste}, which currently implements vertex and cell-centre models. For now we consider a range of different examples, where one particular approach was typically chosen.

Intestinal crypt mechanics have been widely studied with IBM models. Meineke et al. \cite{meineke2001cell} developed the first lattice-free IBM model of cell motion in the intestinal crypt; the crypt is modelled as a cylindrical surface and cell centres interact via linear spring forces. The model was developed to overcome limitations of earlier, lattice-based CA models of the intestinal crypt such as \cite{loeffler1986intestinal}. In particular, the model \cite{meineke2001cell} allows for continuous time and space steps (to computational resolution), avoids unrealistic shifting of columns of cells, and cell motion is not based on rules involving age or knowledge of neighbouring cell types, in contrast to \cite{loeffler1986intestinal}. Furthermore, the model can be adapted to determine how the cell dynamics change when the idealised cylindrical geometry of the crypt is replaced by a more realistic test-tube shaped geometry.

More recently, other authors have included more biological detail into IBM models of the intestinal crypt~\cite{vanleeuwen2009integrative, buske2011comprehensive, fletcher2012mathematical,mirams2012theoretical, baker2014quantification}. 
For example, van Leeuwen et al.~\cite{vanleeuwen2009integrative} construct a multiscale model of the intestinal crypt which couples dynamic, subcellular models of the cell-cycle and Wnt signalling pathway with a cell-scale mechanical model based on Meineke et al.'s \cite{meineke2001cell}. Their model simulations reveal that relating cell proliferation explicitly to the cells' biochemical environment or `niche' more naturally explains the experimental data than the idea that cells are intrinsically `stem' or `non-stem' cells independent of their environment. Building on the van Leeuwen et al. model, Mirams et al. \cite{mirams2012theoretical} studied how monoclonal conversion in the colonic crypt (i.e. when all cells in a crypt have emerged from the same stem cell) is affected by mutations causing changes in the proliferative and adhesive properties of cells, as well as the spatial location at which these mutations occur. A particularly robust finding of theirs was that mutations must occur in the bottom few locations of the crypt in order to have any chance of their descendants taking over the crypt. Thus a better understanding of spatial dynamics can lead to a better understanding of the transitions from healthy to diseased states.

A notable early application of the subcellular element method was to individual cell rheology \cite{sandersius2008modeling}; here the authors found evidence that they need to incorporate active cytoskeletal rearrangement to account for long-time/large-strain behaviour. The incorporation of active processes was subsequently investigated by the same authors in both a cellular and multicellular context \cite{sandersius2011emergent}. They give favourable comparisons to experiments of a range of different types - cell stretching under large-strain conditions, tissue fluidity in embryos and streaming patterns in collective cell motion (see also \cite{sandersius2011chemotactic}). Other applications of this method to epithelia include those by Nie and co-workers \cite{christley2010integrative,gord2014computational}, who applied the subcellular element method, giving particular attention to its implementation on graphics processing units (GPUs), to investigate the epidermal epithelium, finding for example that a polarized distribution of adhesion proteins, which inhibits cell detachment, in combination with asymmetric cell divisions provides a robust mechanism for the formation of the stratified structure of skin layers. The subcellular element method is one of the most detailed frameworks available for modelling epithelial dynamics and will likely continue to be applied to a range of systems.


\section{Continuum models} \label{sec:continuum_models}

\subsection{Overview and motivation}

Biological tissues such as epithelia have been modelled as continua for almost a century. 
Macroscale representations of growth have been of particular interest, stimulated initially by D'Arcy Thompson's~\cite{thompson1917on} simple geometric theories. 
Recent reviews of continuum theories for growth and remodelling, and on continuum and discrete representations of growth, can be found in~\cite{ambrosi2011perspectives} and~\cite{jones2012modeling}, respectively. 

Simple continuum models, based on conservation principles and constitutive assumptions (e.g. Darcy's law for cell movement through a fluid), have been developed to describe epithelial tissues including the intestinal crypt. Such models are reviewed and compared to cell-based models in~\cite{osborne2010hybrid, walter2009comparison}, in the context of cancer in the colorectal crypt epithelium. 
Simple convection-diffusion-reaction models have recently been used to study the effects of chemical signalling on crypt formation in the intestine \cite{zhang2012reaction}, tissue stratification in epithelia \cite{chou2010spatial,ovadia2013stem} and for formulating an inverse problem for proliferation rates in the crypt \cite{figueiredo2013physiologic}. 

It is often difficult, however, to relate mechanisms included in the continuum models to mesoscopic processes such as those involved in individual cellular rearrangements. In this section we primarily study epithelial relaxation and rearrangement by taking continuum limits of some of the IBMs discussed above, before returning briefly to recent developments in continuum modelling directly incorporating more mesoscopic information.


\subsection{Relations between discrete and continuum modelling}

Murray et al. \cite{murray2012classifying} argue that nonlinear diffusion equations may represent a conceptual continuum framework for comparing different discrete models of epithelial cell movement, and also for `inverting' discrete to continuum derivations in order to formulate discrete models that are compatible with a particular nonlinear diffusion equation. In particular, they note \cite{murray2012classifying} that continuum limits of IBM models frequently result in equations of the form, here in one-dimension,

\begin{align}
\frac{\partial n}{\partial t} = \frac{\partial}{\partial x}\left(D(n)\frac{\partial n}{\partial x}\right),
\label{eq:diff}
\end{align}
where $n(x,t)$ is the cellular density at position $x$ and time $t$ and $D(n$) is a nonlinear diffusion coefficient. Here we have neglected growth, and just consider the relaxation of tissues towards equilibrium; we return to this issue in the discussion. A benefit of deriving such a coarse-grained description is the availability of mathematical analysis tools for the study of such models. Before we discuss this further, we consider some examples.

Murray et al. \cite{murray2009discrete} consider a discrete model of one-dimensional cell motion, in the overdamped limit for which viscous forces are dominant and similar in form to the cell-centre model in \cite{meineke2001cell}. By converting the discrete cell index into a continuous index and assuming that cell numbers are large, they obtain an equation of the form (\ref{eq:diff}), with 

\begin{align}
D(n)= \frac{k}{\eta n^2}
\label{eq:murray}
\end{align}
where, as above, $n$ represents the cell density, $k$ is the spring constant associated with cell-cell interactions and $\eta$ is a viscous drag coefficient characterising resistance to cell motion. Murray et al. \cite{murray2009discrete} demonstrate good agreement between simulations of the discrete and continuum models and note that models of this sort have appeared in the context of `fast' diffusion phenomena.


Using an alternative approach, Lushnikov et al. \cite{lushnikov2008macroscopic} consider continuum limits for a (lattice-based) cellular Potts model of collective cell movement. As discussed above, interactions are incorporated via an energetic (Hamiltonian) approach; for example, cell-cell adherence, interaction with the extracellular matrix and an excluded volume constraint (ensuring cells do not overlap) are all included in the `effective energy'. They take the surrounding medium to be viscous and neglect inertial effects. They first derive a chemical master equation before taking a continuum limit. Despite their different approach and model type, this leads to a nonlinear diffusion equation of the same form as (\ref{eq:diff}), this time with
\begin{align}
D(n) = C\frac{1+(\frac{n}{n_0})^2}{(1-\frac{n}{n_0})^2}
\label{eq:alber}
\end{align}
where $n$ is the cell density, $C$ is a constant and $n_0$ is the equilibrium cell density, equal to one over the local cell rest length (see also \cite{murray2012classifying,markham2014modelling}). 


As noted above, Murray et al. \cite{murray2012classifying} explain how different discrete models can be compared by analysing the functional form of the nonlinear diffusion coefficient that is generated when a continuum limit of the cell-based equations is taken. 
As further support for this approach, they found that they could derive an equivalent Newtonian force law at the cell level which recovers the same continuum description of the CPM as Lushknikov et al. \cite{lushnikov2008macroscopic}. 
This illustrates how information at the continuum level can be used to infer information about possible behaviour at the cell scale and to compare alternative cell-based descriptions. Furthermore continuum models are generally much easier to analyse than cell-based models; they can be used either independently or in combination with cell-level simulations. For example, Murray et al. \cite{murray2011comparing} found, by adding a cell proliferation term to equation (\ref{eq:diff}) with diffusion coefficient (\ref{eq:murray}), that a bifurcation in model behaviour occurs - if the diffusion is too slow relative to cell division rate, the system does not reach a steady state but instead `blows up'. Since this is biologically unrealistic under normal conditions, this aids in understanding the important balances of different biological processes, as well as providing guidance for parameterisation of the IBM model. 

There are several directions that require further work, however. It is not clear, in particular, how growth, topological changes, subcellular signalling and energy transduction should be rigorously incorporated. Extensions of these derivations to higher dimensions also remain to be established. While Fozard et al. \cite{fozard2010continuum} also investigate continuum limits of a discrete mechanical model of cell motion, systematically considering both slowly varying and heterogeneous cell parameter limits, they also limit their analysis to one dimension. In addition, the approaches of Murray et al. \cite{murray2012classifying,murray2011comparing} and Fozard et al. \cite{fozard2010continuum} are based on limits of deterministic models; it is likely valuable to retain more explicit considerations of stochastic effects, as in Lushknikov et al.'s approach \cite{lushnikov2008macroscopic}. Middleton et al. \cite{middleton2014continuum}, building on the work of Newman and Grima \cite{newman2004many}, present a similar approach for a stochastically-forced point-mass cell-centre model (i.e. a Langevin equation model) of cell migration in the presence of adhesive effects, taking careful account of the correlations induced by strong mechanical interactions between cells. In principle this approach is applicable in three-dimensions but only one-dimensional comparisons are carried out.

Contrasting with the above `bottom-up' approach, we briefly mention the more `top-down' generalised continuum theories, which may provide one alternative or, more appropriately, a complement to `bottom-up' modelling \cite{jones2012modeling,ambrosi2011perspectives}. The sophistication of continuum models has also increased significantly in parallel with discrete models, with the introduction of additional mesoscopic state variables, multiple phases and extensions of continuum kinematics. These generalised continuum theories are an important approach to representing phenomena such as growth, rearrangements, plastic deformations and other dissipative phenomena, as well as coupling to processes such as chemical signalling and mechanotransduction. Here we can only briefly review this large topic and refer the reader to the books \cite{ziegler1983introduction,maugin1999thermomechanics,houlsby2007principles,gurtin2000configurational,gurtin2010mechanics} for excellent introductions to such generalised continuum theories (here ultilising subtle concepts such as `internal state variables' and `configurational forces'), and the articles \cite{garikipati2009kinematics,jones2012modeling,menzel2012frontiers,ambrosi2011perspectives} reviewing, among other topics, multiphase theories, continuum thermodynamics and the so-called multiplicative decomposition of the continuum deformation gradient in the context of growth and remodelling of biological materials.

This approach can help bypass the technical difficulties of homogenisation of underlying discrete models; however it is probably best combined with partial homogenisation, statistical mechanics and simulation targeted at modelling particular state variables such as the `growth tensor' often used in continuum biomechanics \cite{jones2012modeling,ambrosi2011perspectives}. This filling in of mesoscopic details, providing appropriate interpretations and determining the validity of a continuous description is the central difficulty involved in applying these generalised continuum models . An interesting practical example of tackling this challenge is \cite{blanchard2009tissue} (see also \cite{blanchard2011measuring}) where the kinematic tools of continuum field theory were used, along with statistical fitting methods, to quantify deformation rates at both the tissue and mesoscopic cellular level (cell shape changes, sliding and rearrangements) in epithelial tissues to assess their impact on development. The rearrangement/intercalation deformation tensor they define provides an example of a mesoscopic `microstructural' variable that fits well with the approach of generalised continuum theories. Similarly, Bonnet et al. \cite{bonnet2012mechanical} investigated the applicability of continuum theory to epithelia through the use of laser ablation experiments, live imaging and statistical fitting methods, finding that the continuum description was well justified (see also their related earlier work on connecting statistical processing of foam structures to continuum theories \cite{graner2008discrete,marmottant2008discrete,raufaste2010discrete}).  


\section{Discussion} \label{sec:discussion}
As detailed in this review, epithelial tissue provides an important experimental model system for investigating a broad range of intertwined biological phenomena, including cell lineage dynamics, tissue repair, tissue homeostasis, tissue development and tumorigenesis. Many features of these processes are still not fully understood, however, and in particular we face three key challenges, raised in the introduction: I) choosing which of the diverse biological processes, spanning multiple spatial and temporal scales, to include in our mathematical models (\textit{scale and resolution of models}); II) how to best compare models of different types (\textit{model-model comparison}) and III) how to best connect our models to data (\textit{model-data comparison}).

We first summarise the features of the various modelling approaches considered in this review, and then discuss some possible approaches for tackling the above challenges more systematically.


\subsection{Summary of modelling approaches considered}

Compartmental models provide a robust and high-level approach suitable for investigating feedbacks and signalling, while having limitations on the resolution of spatial and mechanical factors. Individual-based models allow for higher-resolution descriptions of cellular motion and mechanics. As such they offer many new insights but can be difficult to analyse and are still subject at present to computational limitations for large simulations. Continuum models provide an approach with a spatial and mechanistic resolution lying between broad-brush compartment models and detailed discrete models, and standard formulations typically lack the detail to permit comparisons with important cellular-level data. While each approach has its strengths, and in many cases each can be used independently of the other, there is much to be gained from methods allowing the comparison of different models. Continuum limits of discrete models are one approach to tackling this; alternatively, modern continuum models tend to include much more `microstructural' information than they previously did. This apparent `meeting in the middle' offers many opportunities to incorporate and combine the strengths of different approaches, but requires practitioners to first be aware of the range of alternatives - one aim of this review is to help further this goal.


\subsection{Future directions - inference}
Addressing questions about how much detail to include, which model class(es) to use and how to relate models to each other and to real biological phenomena remains an art rather than science. These questions have typically been tackled in an \textit{ad-hoc} manner in mathematical biology; on the other hand, advances in computation, formalisation of inference methods and new imaging methods should enable some aspects of these questions to be addressed more systematically. 

\begin{figure}[hbt]
\centering
\includegraphics[width=.8\linewidth]{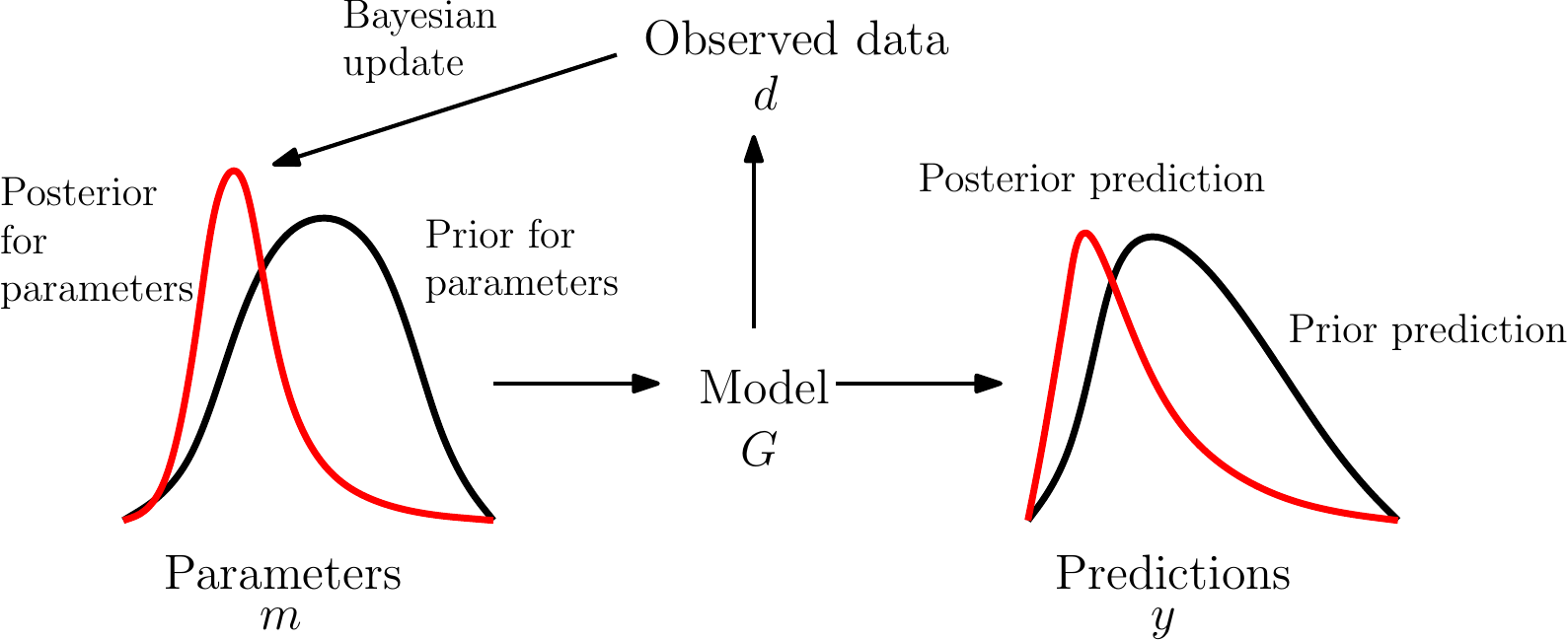}
\caption{In the Bayesian approach one starts from a fully-specified probabilistic model, which can generate predictions based on assumptions about model parameters which are expressed in terms of a \textit{prior} probability distribution. If new data are obtained, one can use the Bayesian conditioning procedure to update the information on model parameters into the form of a \textit{posterior} probability distribution. This can then be used to generate new predictions or carry out model checks, and this process can be repeated if further data are obtained. Based on \cite{iglesias2014}.}
\label{fig:bayes}
\end{figure}

In particular, in statistics, the rise in computational power has seen a resurgence of interest in Bayesian inference methods, typically through Markov chain Monte Carlo sampling methods (see for example the account in \cite{berger2000bayesian}). One of the biggest strengths of the Bayesian approach is that it provides a formalised inference framework for combining models with disparate data sources and other information and for characterising the uncertainty and/or `underdetermination' in parameters \cite{gelman2013bayesian,tarantola2005inverse}. This provides one possible inference-based perspective for assessing models and relating them to data\footnote{There are of course many sophisticated `frequentist' approaches, but we do not consider these here.}. The key here is to interpret probabilistically \textit{both} observable data and model parameters (this latter part being particular to the Bayesian inference approach), which allows/forces one to define a `full probability model' or joint distribution for observable data and model parameters \cite{gelman2013bayesian,tarantola2005inverse}. By conditioning on a sample of observed data, one can update \textit{prior} estimates of model parameters (expressed in terms of probability distributions) and as-yet unobserved data to obtain a new (\textit{posterior}) predictive model. One appealing aspect of this approach is that this posterior can then be used either as a prior to be updated on the basis of further data or as a predictive model, which can also be subject to model checking/tests\footnote{We note that parameter estimation, uncertainty quantification (probabilistic predictions from uncertain input) and model checking can all be carried out within the Bayesian framework, as nicely described for example in \cite{gelman2013bayesian,gelman2013philosophy}. We also note that, as argued by these authors, while it may be difficult to obtain good prior information about some parameters, this approach forces one to explicitly acknowledge and (at least formally) account for this uncertainty. See also the high-level overview \cite{iglesias2014} and more mathematically-detailed \cite{stuart2010inverse,dashti2014bayesian} linking Bayesian methods to the field of uncertainty quantification.}. There are of course a number of mathematical and modelling complexities, especially when dealing with the high or even infinite-dimensional nature of parameter estimation and uncertainty quantification for truly spatio-temporal models. This represents an exciting interface between applied mathematics and statistics; we refer the reader to \cite{stuart2010inverse,dashti2014bayesian} for more mathematical details in the context of PDE models. We illustrate the general Bayesian approach in Figure \ref{fig:bayes}. 

The recent work of Oden et al. \cite{oden2010general,oden2013selection,hawkins2013bayesian} illustrates this approach in the context of tumour modelling. They combine a modelling framework of appropriate (spatio-temporal) resolution - the generalised phenomenological framework of continuum thermodynamics of mixtures - with the formalised methods of Bayesian inference and uncertainty quantification as outlined above. Their approach is still rather general and we refer the reader to this work for more information. Of particular note, however, is that even a deterministic continuum model such as that developed in \cite{oden2010general,oden2013selection,hawkins2013bayesian} is now interpreted stochastically, here as a set of stochastic partial differential equations. With this in mind, we speculate that this `inference-centric' approach may further bridge the gap between model types, with both initially discrete or continuous and deterministic or stochastic models ultimately having common interpretations in terms of probabilistic models linking both observable data and model parameters. By translating models into this common language\footnote{One well-known advocate of an approach such as this was the physicist E.T. Jaynes, who (re-)interpreted a number of solution methods such as the methods of statistical mechanics and regularization methods in inverse problem theory as problems of (Bayesian) inference (see e.g. \cite{jaynes2003probability,jaynes1989jaynes}). Similar influential views may be found in the work of geophysicist A. Tarantola, e.g. \cite{tarantola2005inverse,mosegaard2002probabilistic}, along with some proposed improvements to aspects of the traditional Bayesian approach.} one could, in principle, facilitate model comparisons and derivations of one type from another (both `bottom-up' and `top-down').

Mathematical biology is an area ripe for application of methods such as these - not just to the newest or most complicated models, though these are certainly of major interest, but also to return to previous models or theories that often far outstripped available data. This will should enable us to systematically test, validate and improve on these models, starting from good empirical information. 


\subsection{Final thoughts}

In general, advances in imaging, computation and statistical methods have brought models and data closer together and this means that we can develop better ways of testing, comparing and choosing between different modelling approaches. Methods are needed for organising and summarising, in an understandable way, different models, interesting data features and complex couplings of biological processes. In this discussion we have speculated that treating specific problems of mathematical biology as problems of inference provides a promising guide to choosing and comparing appropriate models, characterising uncertainty and even deriving one model type from another. 

This perspective is but one promising approach to be further developed and employed. Furthermore, as we have noted with earlier models, modeller creativity will often outstrip available data; in the right balance this is not necessarily a bad thing. While we feel that some of the most exciting challenges presently involve directly relating models and data, novel models will likely continue to provide conceptual foundations extending even further into the future.


\section*{Acknowledgments} 
O.J.M. is funded by the Biotechnology and Biological Sciences Research Council through grant BB/K017578/1.
A.G.F. is funded by the Engineering and Physical Sciences Research Council and Microsoft Research, Cambridge through grant EP/I017909/1.


\bibliographystyle{plain}


\end{document}